# NEURAL (EEG) RESPONSE DURING CREATION AND APPRECIATION: A NOVEL STUDY WITH HINDUSTANI RAGA MUSIC

Archi Banerjee[1,2], Shankha Sanyal[1,2], , Souparno Roy[1,2], Sourya Sengupta[3],

Sayan Biswas[3], Sayan Nag[3*], Ranjan Sengupta[1] and Dipak Ghosh[1]

[1]Sir C.V. Raman Centre for Physics and Music, Jadavpur University

[2]Department of Physics, Jadavpur University

[3]Department of Electrical Engineering, Jadavpur University

* Corresponding Author

## ABSTRACT

*What happens inside the performer's brain when he is performing and composing a particular raga? Are there some specific regions in brain which are activated when an artist is creating or imaging a raga in his brain? Do the regions remain the same when the artist is listening to the same raga sung by him? These are the questions that perplexed neuroscientists for a long time. In this study we strive to answer these questions by using latest state-of-the-art techniques to assess brain response. An EEG experiment was conducted for two eminent performers of Indian classical music, when they mentally created the imagery of a raga Jay Jayanti in their mind, as well as when they listened to the same raga. The beauty of Hindustani music lies in the fact that the musician is himself the composer and recreates the imagery of the raga in his mind while performing, hence the scope of creative improvisations are immense. The alpha and theta frequency rhythms were segregated from each of the time series data and analyzed using robust non MFDXA technique to quantitatively assess the degree of cross-correlation of each EEG frequency rhythm in different combination of electrodes from frontal, occipital and temporal lobes. A strong response was found in the occipital and fronto-occipital region during mental improvisation of the raga, which is an interesting revelation of this study. Strong retentive features were obtained in regard to both alpha and theta rhythms in musical listening in the fronto-temporal and occipital-temporal region while the features were almost absent in the thinking part. Further, other specific regions have been identified separately for the two separate conditions in which the correlations among the different lobes were the strongest.*

---

*nagsayan112358@gmail.com

# INTRODUCTION

Creativity in musical performances is gaining rapid importance as a field of research in recent years; the primary question being how to assess creative correlates during a performance. The primary hitch in these approaches being the proper definition of creativity. A general definition is given in [1] which says creativity is the production of something both novel and useful. The literature on musical creativity is quite large with its facets ranging from musicology [2], psychology [3, 4], cognitive science [5] and art history [6]; but it is the cognitive modalities of creative improvisation during a performance that is currently drawing huge interest.

Musical improvisation by far the most challenging task that an artist has to undertake, requiring the real-time generation and production of novel melodic and rhythmic sequences in line with an ongoing musical performance. Thus, understanding musical improvisation is crucial to understand how in general creative processes are conducted in human being. A recent neuro-imaging study [7] reports increased surface area for subjects reporting high levels of musical creativity which suggests that domain-specific musical expertise, default-mode cognitive processing style, and intensity of emotional experience might all coordinate to motivate and facilitate the drive to create music. Earlier brain-imaging studies pointed the importance of Pre-Frontal cortex (PFC) in case of creative thinking [8-9]; while study on Jazz musicians [10] reveal that improvisation (compared to production of over-learned musical sequences) was consistently characterized by a dissociated pattern of activity in the prefrontal cortex: extensive deactivation of dorsolateral prefrontal and lateral orbital regions with focal activation of the medial prefrontal (frontal polar) cortex. In Western music, a number of neuro-imaging studies have been conducted on Jazz musicians and pianists [10-13]; where activity observed in the PFC included both deactivation of the DLPFC and lateral orbital (LOFC) regions and focal activation of the medial prefrontal cortex (MPFC). Also [10] reveals a state of free-flowing complex musical ideas that may result from the combination of internally generated self-expression (via the MPFC) and attenuation of activity in the DLPFC.

But, what about musical improvisation in Hindustani music (HM)? Till date, no study tries to harvest the immense potential that a Hindustani musician has to offer when it comes to the study of creativity in musical performances since there is no notation in HM system like western music and the musician is himself the composer. A musician while performing expresses the *raga* according to his mood. Thus there are differences from one rendition to another. Even if an artist sings or play same *Raga* and same *Bandish* twice then there is supposed to be some dissimilarity in between the two performances. These differences in the rendition of a raga several times on different days are generally called improvisation. Unlike symphony or a concerto, *Raga* is unpredictable; it is eternally blooming, blossoming out into new and vivid forms during each and every performance which is the essence of "improvisation" [14]. The performers of Hindustani raga music insist that while performing or composing a musical piece, they have a visual imagery of that particular composition in their mind which helps them to improvise and reach to the audience better. The literature regarding perception and imagination of a musical stimuli involving Hindustani *raga* music is quite scarce, though it is quite rich and diverse when it comes to the variety of emotions induced by it [15-18]. Simply put, a *raga* in Hindustani Classical music is a musical theme created by choosing a specific set of notes from within an octave. In the broadest sense, the word '*raga*' refers to 'color', more specifically the emotion or mood produced by a particular sequence or combination of pitches [19]. Different sets of notes evoke different moods and inspire different feelings [20]. Issues of artistic creativity might involve too many variable parameters such as personality, inherent artistic ability, mood etc and at the onset it may be difficult to tackle all these. But, with the onset of neurocognitive science we have robust neuro-imaging techniques with which we look forward to have an accurate insight of brain response while an artist is mentally creating as well as listening to a particular musical composition (*raga* in our case). The neuro-dynamical effect of mentally composing as well as listening to a '*raga*' was studied in this work taking the help of an experienced performer of Hindustani music.

The Electroencephalography (EEG) is a neuro scientific tool which gives precise measurement of the timing of the response because of its high temporal resolution, which is an advantage of EEG over PET (Positron Emission Tomography) or fMRI (functional Magnetic Resonance Imaging). Assessment of creative thinking using EEG based studies has been one of the predominant issues of neuroscience that provided contradictory results in the past. A group of researchers in favor of the lateral dominance have, come to the conclusion that the right hemisphere and its regions are specialized for creative tasking [21-24], while a number of studies have come up with contradictory evidence [25, 26]. Although, it seems a bit impractical to assume that only the right hemisphere is involved in the process of creative thinking as there is stupendous amount of correlation among the hemispheres involved in different perceptual and cognitive tasks [27]. Most of these works make use of coherence properties between the lobes using linear power spectral analysis in various frequency ranges to assess the amount of interdependence. Coherence, a parameter obtained from spectral analysis of the EEG, is the normalized cross-spectrum of two signals and reflects the correlation between them with respect to frequency. Applied to EEG analysis, the value of coherence lies in its providing data on the relationships between the electric oscillations recorded from two locations on the skull [28]. A number of EEG studies have been conducted with the help of power spectral analysis to assess musical emotions [30-33]. These studies mostly speak in favor of the lateralization theory when it comes to the processing of positive and negative emotions. Most of the power spectral study show that activity in the alpha frequency band is negatively related to the activity of the cortex, such that larger alpha frequency values are related to lower activity in the cortical areas of the brain, while lower alpha frequencies are associated with higher activity in the cortical areas [31, 33]. It has also been shown that pleasant music would elicit an increase of Fm theta power [32]. In this work we sought to find the degree of cross-correlation between different lobes in alpha and theta EEG bands for an experienced Hindustani performer while he is mentally improvising as well as listening to a Hindustani raga. No previous study, to our knowledge has studied the non-linear aspects of EEG to study creative musical imagery on trained musicians with auditory stimuli (in our case – a sample of Hindustani *raga* music).

The brain is said to be the most complex structure found in human body and the EEG signals generated from brain are essentially non-stationery and scale varying in nature [34, 35]. There has been increasing evidence that spontaneous brain responses, as recorded from single neuron to millions of neurons, is not necessarily random; on the contrary, it could be better characterized by persistent long-range temporal correlations and scale-free dynamics[36, 37]. Different scaling exponents can be revealed for many interwoven fractal subsets of the time series. So, a multifractal analysis of the data would be more appropriate than a single scaling exponent as is obtained from Detrended Fluctuation Analysis (DFA) [38].

The multifractals are fundamentally more complex and inhomogeneous than monofractals and describe time series featured by very irregular dynamics, with sudden and intense bursts of high-frequency fluctuations. The multifractal detrended fluctuation analysis (MFDFA) was first conceived by Kantelhardt et al. [39] as a generalization of the standard DFA. MFDFA is capable of determining multifractal scaling behavior of non-stationary time series. The non stationary time signals generated from different electrodes are best analyzed with the MFDFA technique [40]. Cross-correlation and Coherence are the measures which are generally used to measure EEG signals obtained from two different regions of brain [41], but these are based on assumption that the signals are linear, which they are not. A few studies made use of the non linear complex nature of brain [42-44], but again focused on specific frequency bands. We analyzed EEG signals appearing from two different groups of electrodes appearing in three different lobes of the brain using a technique called Multifractal Detrended Cross correlation Analysis (MFDXA) [45], which can unveil the multifractal features of two cross-correlated signals and higher-dimensional multifractal measures and can provide a quantitative parameter depicting 'degree of cross-correlation'. The strength of the MF-DXA method is seen in various phenomena using one- and two-dimensional binomial measures, multifractal random walks (MRWs) and financial markets [46-48]. MFDXA technique has also been successfully applied to in the analysis of musical improvisation from a musical signal [49].

In this work, we applied the MFDXA technique to assess the inter lobe as well as intra lobe alpha and theta cross correlation from EEG recordings of the same person during musical imagery and perception. It is worth mentioning here, that though a number of previous works referred here rely on fMRI or PET data, robust non-linear techniques such as MFDXA can only be applied on EEG time series data obtained from different electrodes in the brain. To serve our purpose, we took EEG data from two eminent Hindustani music performers (one vocalist and the other instrumentalist) having an experience of more than 30 years in practicing Hindustani *raga* music. The subjects chose to think of the raga *'Jay Jayanti'* in their mind; they created the imagery of the *aalap* of the raga in his mind. Next, they perceived the *alaap* of the same raga sung/played by themselves after a 5 min relaxation period. We identified three different lobes of the brain namely frontal, temporal and occipital whose functions concur with our work. The frontal lobe is associated with reasoning, cognitive processing and expressive language, the temporal lobe is important for interpreting sounds and the language we hear while the occipital lobe is important for interpreting visual stimuli and information processing. The frontal lobe plays a key role in emotional processing, appraisal of musical stimuli and other higher order cognitive skills [28-32]. We therefore chose two pair of electrodes from each of these lobes (F3/F4 from frontal, T3/T4 from temporal O1/O2 from occipital) to study the brain electrical response of the artist while he is creating as well as listening to a raga of his choice. The signals from the two different groups of electrodes were separated using Wavelet Transform (WT) [50] technique into alpha and theta frequency rhythms and analyzed with the help of MFDXA technique. The resultant cross correlation exponent corresponding to each frequency rhythm gives the degree or the amount by which the two signals are correlated. When a musician is listening and also imaging a certain raga in their mind, they claim that the role of musical expectancy as well as the memory of the just listened phrases and the possible connection to this with immediately following expected musical events is higher. This may lead to strong correlations of specific frequency bands in the lobes where the processing of musical imagery and perception takes place. The complexity of the particular lobe may also be suitably affected when the processing of musical imagery or perception is taking place. This was the main objective behind this work, where we quantitatively analyze the arousal based effects in each lobe during creative composition as well as perception of a musical piece, and also analyze the degree of cross-correlation between different lobes of the brain in the alpha and theta EEG rhythms during these experimental conditions. What we look forward to in this paper is to conjure a paradigm in which we can identify the lobes of brain mostly involved during perception and mental improvisation of a *raga* piece.

# MATERIALS AND METHODS
## A. SUBJECTS SUMMARY:

Two male professional performers of Hindustani music (age between 45~50 years, average body weight ~ 60 kg) voluntarily participated in the study. One of them is a renowned vocalist and the other an eminent *sitar* player, both of them performing in stage for more than 30 years. The two performers were chosen keeping in mind their difference in musical pedagogy which means their mode of musical training is different, so we can assume that their mode of expression and creative improvisation is different. Both the subjects are researchers associated with Sir C.V. Raman Centre for Physics and Music, Jadavpur University, Kolkata, India. Non musicians were not involved in this study as creative imagery of a Hindustani *raga* can only be perceived by an experienced performer. The experiments were performed at the Sir C.V. Raman Centre for Physics and Music, Jadavpur University, Kolkata. The experiment was conducted in the afternoon with a normal diet in a normally conditioned room sitting on a comfortable chair and performed as per the guidelines of the Institutional Ethics Committee. All subjects gave written consent before participating in the study, approved by the Ethics Committee of Jadavpur University, Kolkata.

## B. EXPERIMENTAL DETAILS:

Both the musicians (henceforth referred to as Subject 1 and Subject 2) chose the *raga Jay Jayanti* for this particular study. Following the prescription of the performers, both of them were asked to bring a 3 min recording of their own recital of the same music sample (a 3 min *alaap* of *raga Jay Jayanti)*. The *alaap* portion of each raga was chosen as the form of the entire *raga* is essentially established and identified in this part. The *alaap* part gradually reveals the mood of the raga using all the notes used in that particular raga and allowed transitions between them with proper distribution over time. It reflects the temperament, creativity and uniqueness of musical training of a musician. Each of these sound signals was digitized at the sample rate of 44.1 KHZ, 16 bit resolution and in a mono channel. Both the signals were normalized to within 0dB and hence intensity or loudness and attack cue are not being considered. A sound system (Logitech R_Z-4 speakers) with high S/N ratio was used in the measurement room for giving music input to the subjects.

### C. EXPERIMENTAL PROTOCOL:

Each subject was prepared with an EEG recording cap with 19 electrodes (Ag/AgCl sintered ring electrodes) placed in the international 10/20 system.

**Fig. 1** depicts the positions of the electrodes. Impedances were checked below 5 kOhms. The ear electrodes A1 and A2 linked together have been used as the reference electrodes. The same reference electrode is used for all the channels. The forehead electrode, FPz has been used as the ground electrode. The EEG recording system (Recorders and Medicare Systems) was operated at 256 samples/s recording on customized software of RMS. The data was band-pass-filtered between 0 and 50 Hz to remove DC drifts.

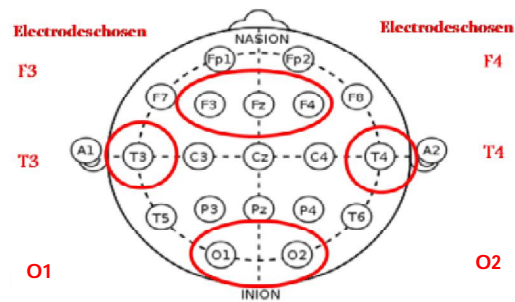

**Fig. 1** The lobes and electrodes chosen for our analysis

Each subject was seated comfortably in a relaxed condition in a chair in a shielded measurement cabin. They were also asked to close their eyes. A sound system (Logitech R_Z-4 speakers) with high S/N ratio was set up in the measurement room that received input from outside the cabin. After initialization, a 20 min recording period was started, and the following protocol was followed:
1. 2min 30 seconds "Before think"
2. 2 min 30 seconds "While thinking raga *Jay Jayanti*"
3. 2 min 30 seconds "After think"
4. 5 min resting period
4. 2 min 30 seconds "Before Listen"
5. 2 min 30 seconds "With listen *raga Jay Jayanti*"
6. 2 min 30 seconds "After Listen"

## METHODOLOGY

We have obtained noise free EEG data for all the electrodes using the EMD technique as in [40] and used this data for further analysis and classification of acoustic stimuli induced EEG features. The amplitude envelope of the alpha (8-13 Hz) and theta (4-7 Hz) frequency ranges were obtained using wavelet transform technique. Data was extracted for these electrodes according to the time period given in the Experimental protocol section i.e. for Experimental conditions 1 to 6.

**A. Wavelet Transform**:
Wavelet transform (WT) forms a general mathematical tool for time-scale signal analysis and decomposition of EEG as well as music signal. We have used WT technique to decompose the EEG and complete music signal into various frequency bands. The DWT [50] analyzes the signal at different frequency bands with different resolutions by decomposing the signal into a coarse approximation and obtains detailed information. The decomposition of the signal into different frequency bands is done by successive high pass and low pass filtering of the time domain signal. In this way time series data of alpha and theta EEG waves as well as desired frequency bands of music signal were obtained corresponding to each experimental condition. On the obtained EEG time series data of alpha and theta rhythms, MFDXA analysis was performed.

B. **Multifractal Detrended Cross Correlation Analysis (MF-DXA)**
We have performed a cross-correlation analysis of correlation between two non-linear EEG frequency signals originating from different lobes of the brain following the prescription of Zhou [45].

$$x_{avg} = 1/N \sum_{i=1}^{N} x(i) \text{ and } y_{avg} = 1/N \sum_{i=1}^{N} y(i) \quad (1)$$

Then we compute the profiles of the underlying data series x(i) and y(i) as

$$X(i) \equiv [\sum_{k=1}^{i} x(k) - x_{avg}] \text{ for } i = 1 \ldots N \quad (2)$$
$$Y(i) \equiv [\sum_{k=1}^{i} x(k) - x_{avg}] \text{ for } i = 1 \ldots N \quad (3)$$

The qth order detrended covariance Fq(s) is obtained after averaging over 2Ns bins.

$$F_q(s) = \{1/2N_s \sum_{v=1}^{2Ns} [F(s,v)]^{q/2}\}^{1/q} \quad (4)$$

where q is an index which can take all possible values except zero because in that case the factor 1/q blows up. The procedure can be repeated by varying the value of s. Fq(s) increases with increase in value of s. If the series is long range power correlated, then Fq(s) will show power law behavior

$$F_q(s) \sim s^{\lambda(q)}.$$

Zhou found that for two time series constructed by binomial measure from p-model, there exists the following relationship [45]:

$$\lambda(q = 2) \approx [h_x(q = 2) + h_y(q = 2)]/2. \quad (5)$$

Podobnik and Stanley have studied this relation when q = 2 for monofractal Autoregressive Fractional Moving Average (ARFIMA) signals and EEG time series [51].

In case of two time series generated by using two uncoupled ARFIMA processes, each of both is autocorrelated, but there is no power-law cross correlation with a specific exponent [10]. According to auto-correlation function given by:

$$C(\tau) = \langle[x(i+\tau) - \langle x\rangle][x(i) - \langle x\rangle]\rangle \sim \tau^{-\gamma}. \quad (6)$$

The cross-correlation function can be written as

$$C_x(\tau) = \langle[x(i+\tau) - \langle x\rangle][y(i) - \langle y\rangle]\rangle \sim \tau^{-\gamma_x} \quad (7)$$

where $\gamma$ and $\gamma_x$ are the auto-correlation and cross-correlation exponents, respectively. Due to the non-stationarities and trends superimposed on the collected data direct calculation of these exponents are usually not recommended rather the reliable method to calculate auto-correlation exponent is the DFA method, namely $\gamma = 2 - 2h(q = 2)$ [52]. Recently, Podobnik et al., have demonstrated the relation between cross-correlation exponent, $\gamma_x$ and scaling exponent $\lambda(q)$ derived from $\gamma_x = 2 - 2\lambda(q = 2)$ [51]. For uncorrelated data, $\gamma_x$ has a value 1 and the lower the value of $\gamma$ and $\gamma_x$ more correlated is the data. In other words, we want to have a quantitative estimate of how the various frequency rhythms originating from different lobes of the brain are correlated when a performer is mentally creating the imagery as well as during the perception of the same raga. The cross correlation coefficient of a particular frequency rhythm is an effective parameter with which we can assess the same and hence have a measure of creativity.

## RESULTS AND DISCUSSION:

Fig. 1 gives the positioning of electrodes according to the 10-20 International system and also marks the three lobes and six electrodes that were chosen for our analysis. The averaged cross correlation coefficient $\gamma_x$ for q = 2 corresponding to the different experimental conditions (for alpha as well as theta EEG rhythms) are computed for different combination of electrodes. As already said, negative values of $\gamma_x$ correspond to strong cross correlation between the two non-linear signals; we report the variation of cross-correlation coefficients in the different lobes while the performers "think" and "listen" to the raga *Jay Jayanti*. What we intend to study is how the degree of cross- correlation between different lobes of brain are affected when a performer is imagining or listening to the same *raga,* so we decided to report the amount of change in degree of cross-correlation in both the cases. The figures that follow (**Fig. 2**) show a quantitative measure of how the cross-correlations between different lobes are affected while musical imagery and listening is constituted in Artist 1's brain. DT (During Think) - BT (Before Think) represent the change in $\gamma_x$ while the performer thinks of the raga; while DL (During Listen) - BL (Before Listen) represent the change in $\gamma_x$ while the performer listens to the same raga and AL/AT represent the After Listen/After Think condition. **Fig. 3** shows how the same combination of lobes behaves when there are no stimuli.

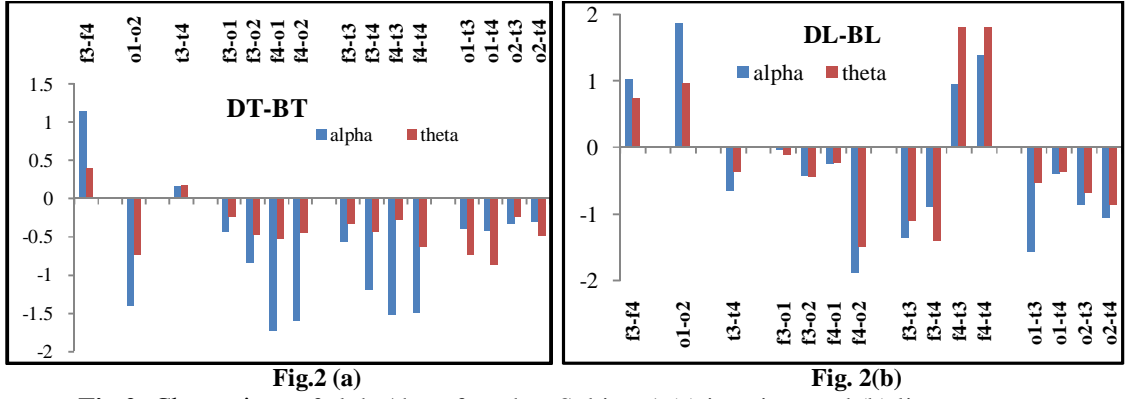

**Fig.2:** Change in γ$_x$ of alpha/theta for when Subject 1 (a) imagines and (b) listens to *raga*

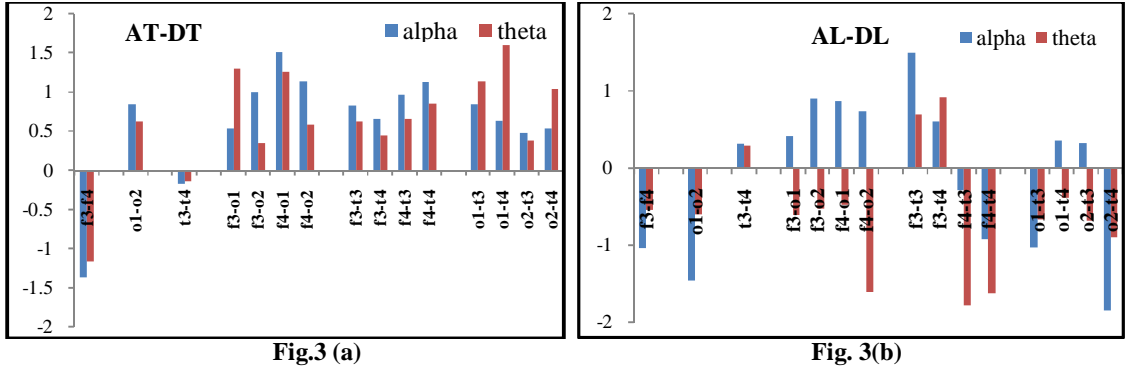

**Fig. 3:** Change in γ$_x$ of alpha/theta after (a) imagination and (b) listening to *raga* by Subject 1

A careful study of the figures shown above reveals the following interesting observations:
1. In the imagination stage, the predominance of alpha correlation is very strong; as we see strong increase in almost all the electrode combinations. The increase in alpha cross-correlations among the fronto-occipital electrode combinations are the strongest followed by the fronto-temporal electrodes and the occipital electrodes. The increase in cross-correlations among the fronto-occipital electrodes may be a signature of the simultaneous interplay of cognitive and visual domain processing in the performer's mind as he conjures up the entire *raga* from his musical memory. It is also interesting to see that the cross-correlation between the fronto-temporal electrodes is also on the rise during imagination period of the *raga;* with the F4-T3 and F4-T4 correlation being affected most significantly. This finding points against the contemporary notion of temporal lobe plays a part only in auditory processing; we see it is being significantly excited even when there are no auditory stimuli. Also the intra-lobe cross correlation for both alpha and theta rhythms in frontal (F3-F4) and temporal (T3-T4) lobe show a decrement while imagining the *raga*.
2. In the listening period, the cross correlation of theta frequency rhythms take a significant part for different electrode combinations along with alpha rhythms. The intra-lobe cross-correlation in frontal (F3-F4) and occipital electrodes (O1-O2) decrease significantly; while the intra lobe cross-correlation in temporal lobe (T3-T4) increases. The inter lobe correlation for F4-O2 electrodes increase consistently for both alpha and theta rhythms. Most significant changes are noticed in the fronto-temporal electrode combinations where F3-T3 and F3-T4 show a significant increase, while the other two register decrease in both the frequency rhythms, though theta is again predominant. In the occipital-temporal domain, the theta cross-correlation in O1-T3 increases significantly while for all other electrode combinations there is an average increase in both alpha and theta cross-correlations. The fronto-temporal and occipital-temporal increase in degree of correlation points in parallel processing of auditory and cognitive information as the artist is constantly relating the *raga* clip with his previous musical expertise and trying to conjure an image of the entire *raga*.

3. In the "after think" part, it is seen that the degree of cross-correlation for both alpha and theta rhythm decreases significantly for all the electrode combinations in different lobes of the performer's brain; with the highest decrease being registered in F4-O1 and O1-T4 combination. This leads to the general conclusion, that after the performer stops creating the *raga* picture in his brain, activity in different regions of the brain gets dilated which essentially leads to the decrease in the degree of cross-correlation in alpha and theta frequency rhythms.

4. In the "after listen" part, even after the removal of stimulus, strong increase in theta and alpha cross-correlations are noticed in almost all the lobes; with the change being maximum in the fronto-temporal combinations as well as intra lobe occipital (O1-O2) and frontal (F3-F4) combinations. In the fronto-temporal combinations, the response is exactly reverse to what is seen in "with stimulus" condition, indicating the neuronal interactions returning to their basal state after the removal of stimulus; while the strong alpha correlations in the other electrode combinations like F4-T4 and O2-T4 indicates signifies the retention of musical auditory memory of the *raga* in the mind of the participants. This retention effect is more pronounced for theta rhythms in all electrode combinations of fronto-occipital and occipital-temporal domain. This retention is almost absent in case of thinking. This retention may have been caused as the subjects probably continued to think/imagine the *raga* even after removal of the auditory stimuli.

The following figures (**Figs. 4 and 5**) show the same response in case of Artist 2:

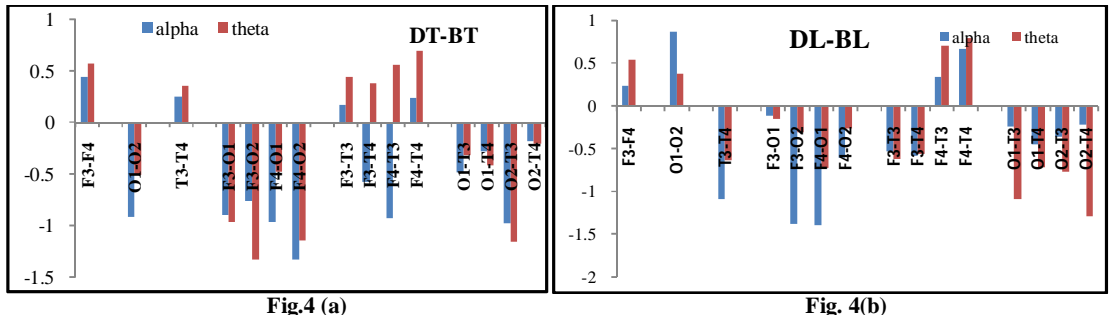

**Fig.4 (a)**   **Fig. 4(b)**

**Fig.4:** Change in $\gamma_x$ of alpha/theta for when Subject 2 (a) imagines and (b) listens to *raga*

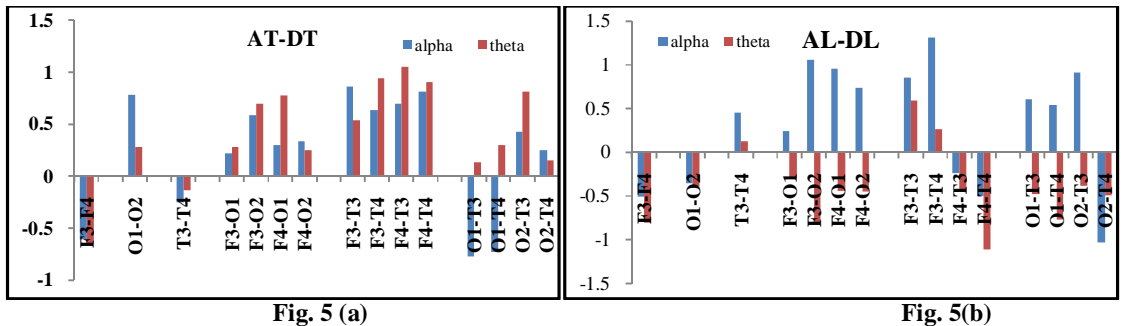

**Fig. 5 (a)**   **Fig. 5(b)**

**Fig. 5:** Change in $\gamma_x$ of alpha/theta after (a) imagination and (b) listening to *raga* by Subject 2

In most cases, the observations are almost similar to the 1st subject, though a few specific deviations from the previous artist are noted; the observations may be summed up as:

1. In the thinking part, the theta cross-correlation also increases significantly for the same electrode combinations as in Subject 1 along with alpha rhythms. Interestingly, for Artist 2, the increase in alpha/theta cross-correlation for the fronto-temporal case is not so pronounced as in the case of Artist 1; with exception being F3-T4 and F4-T3 where the alpha correlation increases significantly while the artist imagines the *raga* in his mind. The occipital-temporal cross-correlations again show consistent increase; depicting again the simultaneous processing of cognitive and musical memory during the mental recreation of the *raga*.

2. In the listening part, the observations are almost the same for the previous artist, with the theta cross-correlations being the most predominant; with significant increase in the occipital-temporal cross correlations. The enhancement in occipital-temporal correlations may be a signature of concurrence of auditory stimuli and musical cognition in the artist's brain. Here the intra lobe temporal cross-correlations in alpha frequency domain also show an increase compared to the 1$^{st}$ artist.

3. In the "after think" part, again the response is similar to 1$^{st}$ subject, with the alpha/theta cross-correlations decreasing in most of the cases; with the exception being alpha cross-correlation in O1-T3 and O1-T4 electrode where residual neuronal activities lead to further increase in cross-correlations. The frontal intra-lobe (F3-F4) alpha and theta correlation also increases further showing enhancement of neuronal interactions even after thinking.

4. In the after listen part, the theta cross-correlation mostly bears the signs of retentive features in various lobes; with the degree of cross-correlation increasing further even after the stimulus is removed. Interestingly the intra-lobe occipital (O1-O2) and frontal (F3-F4) correlation increases for both alpha and theta frequency rhythms.

Thus, with the help of this study we have reported for the first time, the effect on alpha and theta cross-correlations when an artist is performing creative tasks and improvising on a musical piece as well as when he is listening to the same musical composition sung by himself.

These responses may vary for other musical pieces as with the change of the musical content of the piece, the emotion associated with it also changes. *Raga Jai jayanti* is conventionally known to evoke happy emotion in listeners. As both subjects were asked to imagine and listen to the same *raga* so, we can safely compare their results.

## CONCLUSION AND FUTURE WORK:

"*If a person can't read or write, you don't assume that this person is incapable of it, just that he or she hasn't learned how to do it. The same is true of creativity. When people say they're not creative, it's often because they don't know what's involved or how creativity works in practice.*" wrote Sir Ken Robinson in his book *The Element* [54]. In this work, we have tried to visualize with the help of robust scientific methods, a general response of human brain while performing creative task. The Multifractal Detrended Cross-Correlation (MFDX) Analysis is a robust non-linear analysis technique developed recently which gives brilliant insight into the degree of cross-correlation existing in different lobes of human brain. The alpha and theta brain rhythms have been previously related to a number of modalities related to musical emotion processing in earlier studies, but this work for the first time, tries to quantify the correlations existing in alpha and theta rhythms to creative appraisal based task with the help of MFDXA technique.

The work presents the following interesting conclusions:

1. The alpha cross correlation plays the significant role when the performer cogitates the *raga* in his mind; with the increase most prominent in the fronto-occipital and fronto-temporal lobe. This shows that inter-lobe cross-correlation is mostly affected during any creative processes compared to the intra-lobe correlation except in case of occipital lobe which is greately affected during thinking for both subjects. Musical creativity and improvisation thus mostly involves interplay between frontal, temporal and occipital lobes, since the correlation among these are the highest as is seen from our study.

2. In case of musical appreciation, i.e. when an artist is listening to the *raga* sung/played by him, the theta cross-correlation is mostly affected in the fronto-temporal and occipital-temporal domain indicating simultaneous processing of cognitive and auditory data.

3. In case of auditory stimulus, the cross-correlations of alpha as well as theta remain consistently high for some time in certain regions of the brain, even after the removal of stimuli while in the case of thinking, the degree of cross-correlation decreases to its basal value in almost all regions. The degree of cross-correlation remains high mostly in the occipital-temporal domain and also in the intra lobe temporal domain. This indicates the auditory stimulus perturbs the brain states to a greater extent compared to the other stimulus which is manifested in enhanced degrees of cross-correlations in particular regions of brain.

4. Although we analyzed the responses of two musicians separately, hoping that each person has definite cognitive appraisals, we could not find any remarkable discrepancies in the results of the two. This may be due to the musical prodigy of the two; both being trained Hindustani musicians their imagination and improvisation of *raga Jay Jayanti* may be on similar lines.

Finally, we can conclude from what we have explored using EEG rhythms from different electrodes extracting features of functioning of different lobes, their intra and inter-correlation, that a straightforward jacketing of auditory-cognition route may be more empirical – the real scenario is a bit complicated, involving simultaneous processing of musical emotions in a number of different lobes. Some of the features of this interdependence, obtained from the degree of cross correlation of alpha/theta rhythms between two lobes, are revealed for the first time from our new data. Future works in this domain includes EEG data from a greater number of samples as well as a variety of musical cognitive tasks which may reveal higher degree of correlation between other lobes also. Also, comparing the data of musical appreciation of performers with that of naïve listeners is also an interesting piece of work; to see the difference in EEG response pattern of professionals and naïve listeners. A new model of emotion elicited by musical stimuli need to accommodate the findings of this investigation in regard to pronounced cross-correlation obtained in the occipital, frontal and temporal lobes. The obtained data may be of immense importance when it comes to studying the neuro-cognitive basis of creativity and alertness to a certain cognitive function.

## ACKNOWLEDGEMENT:

The first author, AB acknowledges the Department of Science and Technology (DST), Govt. of India for providing (A.20020/11/97-IFD) the DST Inspire Fellowship to pursue this research work. One of the authors, SS acknowledges the West Bengal State Council of Science and Technology (WBSCST), Govt. of West Bengal for providing the S.N. Bose Research Fellowship Award to pursue this research (193/WBSCST/F/0520/14). SR acknowledges the Department of Science and Technology (DST), Govt. of West Bengal for providing Junior Research Fellowship under Research Project Grant No. [860(Sanc.)/ST/P/S&T/4G-3/2013].